# Production of a Titanium Ion Beam Using Fluorides and Fluorine-Forming Precursors in an ECR Ion Source

*D.K. Pugachev*, K.I. Berestov, A.E. Bondarchenko, V.N. Loginov, A.N. Lebedev, E.Yu. Kidanova, K.I. Kuzmenkov, V.E. Mironov, D.S. Podoinikov, D.S. Yakovlev*

*Joint Institute for Nuclear Research, Flerov Laboratory of Nuclear Reactions, Dubna, Moscow Region, 141980, Russia*

**e-mail: pugachev@jinr.ru*

**Abstract.** The production of multiply charged titanium ions is of significant interest for accelerator-based experiments, surface modification technologies, and applications in nuclear physics. Due to the low saturated vapor pressure of titanium at temperatures below 1000 °C, the generation of titanium ion beams from an electron cyclotron resonance ion source (ECRIS) remains a challenging task. The choice of titanium vapor injection methods into the ECR plasma plays a key role in achieving high ionization efficiency, plasma stability, and reliable long-term source operation. In this work, alternative approaches for the injection of neutral titanium atoms into the ECR plasma are investigated, focusing on the use of fluorine-containing compounds. Titanium fluorides ($TiF_3$, $TiF_4$) possess a relatively high saturated vapor pressure at moderate temperatures, allowing to use standard resistively heated ovens operating up to 1000 °C. Another method involves the in-situ formation of titanium fluorides inside the plasma chamber via chemical reactions between metal titanium and the dissociation products of sulfur hexafluoride ($SF_6$). These approaches enable controlled and efficient titanium injection into the ECRIS plasma with satisfactory extracted ion beam stability.

## 1. Introduction

The Super-heavy Element Factory (SHE Factory) at the Joint Institute for Nuclear Research (JINR, Dubna) is designed for experiments on synthesis and investigation of super-heavy elements. The facility is equipped with the modern DC-280 cyclotron, which makes possible to study nuclear reactions with extremely low cross sections. Previously, a large-scale research program using $^{48}Ca$ ion beams was successfully carried out at the U-400 cyclotron of the Flerov Laboratory of Nuclear Reactions, JINR. Several new super-heavy elements were synthesized, the most recent being oganesson with the atomic number of 118. Synthesis of even heavier elements is only possible with new neutron-rich projectiles, such as $^{50}Ti$.

The front-end ECRIS ion source of SHE Factory is requested to produce intense beams of titanium ions with the charge states of (5-11)+ for a few weeks of steady operation and minimized consumption of the expensive isotopically-enriched material. Presently, we routinely use the MIVOC method [1] for injection of titanium atoms into the source. The injected atoms enter the dense parts of the ECRIS plasma, where they are captured after ionization and reach the desired charge state. The extracted ion beam parameters meet the technical requirements, with the $^{50}Ti^{11+}$ currents of up to 5 pμA and material consumption of ~2.5 mg/h. However, request for further increase in the beam intensity is envisaged, which is not possible in the present configuration.

Within the scope of this work, the alternative approach is investigated for injection of titanium atoms into the plasma of the ECR ion source by an evaporation of titanium fluoride powders. These fluorides have a relatively high saturated vapor pressure at moderate temperatures, and it is therefore possible to use for their evaporation a resistive heater with operating temperatures of up to 1000 °C. The second investigated method involves reactive ion etching of the metallic titanium placed inside the plasma chamber. At this, the main plasma in the source is maintained under injection of fluorine containing gas ($SF_6$) into the source, and the volatile titanium fluorides are created directly in the plasma chamber [2] as a result of the chemical interactions between the titanium surface and the impinging fluorine ions and radicals. It is found that both these approaches provide a controllable and efficient supply of titanium atoms into the ECR plasma and maintain satisfactory ion beam stability.

## 2. Ion source

Development of titanium atom injection methods was carried out using the DECRIS-PM ECR ion source [3] installed on the high-voltage platform of the SHE-Factory.The magnetic system of the source is a combination of NdFeB permanent magnets, the axial magnetic field is 1.3 T at the injection, 1.1 T at the extraction, and 0.4 T in the magnetic field minimum. The hexapole magnetic field on the radial wall is 1.1 T. The operating frequency is 14.5 GHz and the injected microwave power is up to 600 W. The inner diameter of the plasma chamber is 70 mm, and the chamber length is 23 cm.

* *corresponding author*

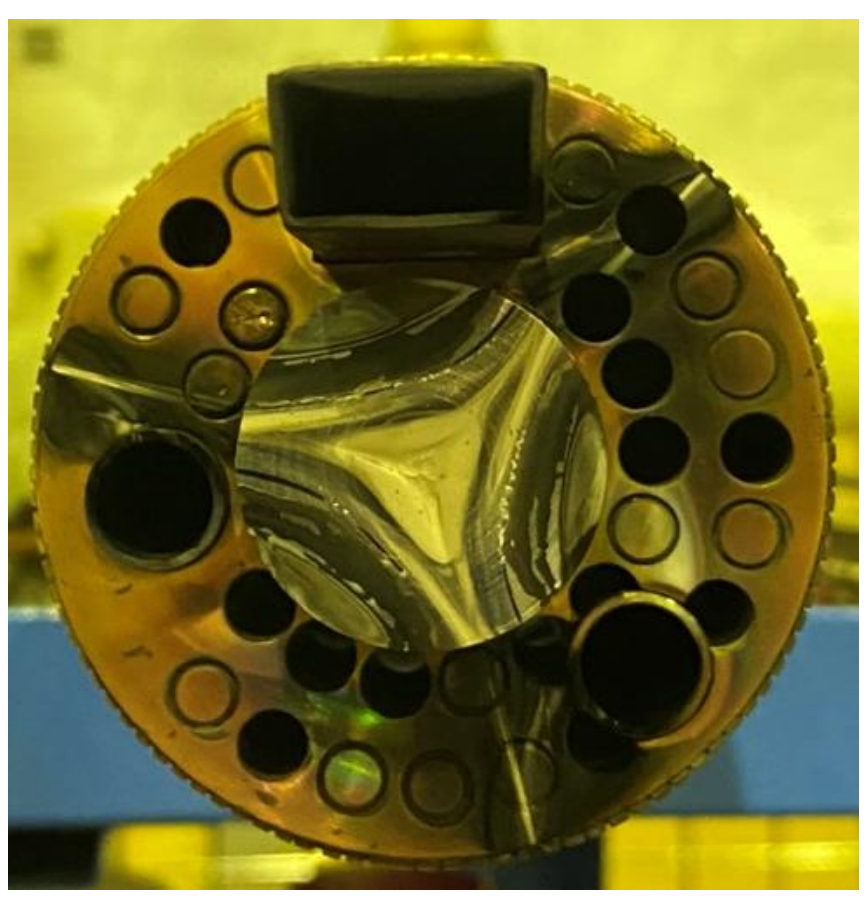

**Fig.1.** – The injection flange of DECRIS-PM source.

The injection flange of the source is shown in Fig.1. Microwaves are injected into the source through the WR-62 waveguide in the Vlasov launcher configuration. The biased electrode with the diameter of 30 mm is placed axially, gas injection is done through two injection ports located such as to avoid a contact with dense plasma flows. When producing the titanium ion beams with MIVOC method, we use both of these ports connected in parallel to the room-temperature vapor transfer line. When needed, the compact resistive oven is placed inside the left port and the support gas is injected through the right port.

With MIVOC operations, various volatile titanium compounds were tested as working materials (e.g., $TiCl_4$, $Ti(C_5H_5)(C_7H_7)$, $(CH_3)_5C_5Ti(CH_3)_3$, etc.) [4,5], which all possess sufficiently high vapor pressure to sustain plasma generation at room temperature: the best results are obtained with $(CH_3)_5C_5Ti(CH_3)_3$. The main limitation of this method is the high sensitivity of the compounds to external factors such as light, moisture, and exposure to air. Failure to maintain the required handling conditions may lead to degradation of the compound. For experiments involving rare isotopes, such as $^{50}Ti$, this issue becomes critical. The maximum intensity of ion beams obtained using the MIVOC method does not exceed 3–5 pμA for ions with medium charge states. This limitation is mainly associated with the high content of impurity elements in the precursor compounds or, for example, in the case of $TiCl_4$, with the relatively large atomic mass of the impurities.

## 3. The titanium fluoride evaporation

Evaporation of metallic titanium using resistive or inductive high-temperature ovens [6] is the successfully tested alternative to MIVOC. In our compact ion source, the method meets with significant technical challenges: evaporation of the material requires a bulky heating system and reliable cooling of the evaporator external elements. Controlling the evaporation process at high temperatures is difficult, non-uniform evaporation and the oven aperture blockage by the condensed material significantly affects the ion beam stability. Still, direct evaporation of titanium allows reaching higher charge states compared to MIVOC by minimizing content of impurities in the plasma and by possibility to use an optimally selected support gas.

Therefore, investigated methods to inject titanium into the source by evaporating the titanium fluorides ($TiF_3$ and $TiF_4$) in a low-temperature oven that is routinely used in our facility for production of calcium ions. The titanium compounds with the natural isotopic composition were used in the test experiments. The oven design is shown in Fig.2a. The stainless-steel crucible with a volume of 0.5 $cm^3$ is heated by mineral-insulated heating cable and the whole assembly is positioned inside the injection plug by linear motion feedthrough. The fluoride powder is placed inside the crucible; the crucible opening is shielded by tantalum wire plug to prevent the material from spilling out. The operating temperature range of the oven is shown in Fig.2b, together with the temperature ranges needed for the fluoride effective evaporation.

The Titanium trifluoride ($TiF_3$) compound is a violet crystalline solid with high thermal stability and with a decomposition temperature exceeding 1000 °C. The saturated vapor pressure of this compound reaches the value required for plasma generation of about $10^{-3}$ Torr at temperatures close to ~800 °C.

Titanium tetrafluoride ($TiF_4$) is a white, hygroscopic solid with significantly higher volatility compared to $TiF_3$. Sublimation of $TiF_4$ begins at 250–300 °C, requesting reduction of the operating temperature of the oven for vapor injection into the plasma of the ECR ion source. The main challenge associated with the use of $TiF_4$ is its high reactivity with moisture and oxygen, necessitating storage in a dry inert atmosphere. Even brief exposure of the material to air leads to hydrolysis and formation of decomposition products.

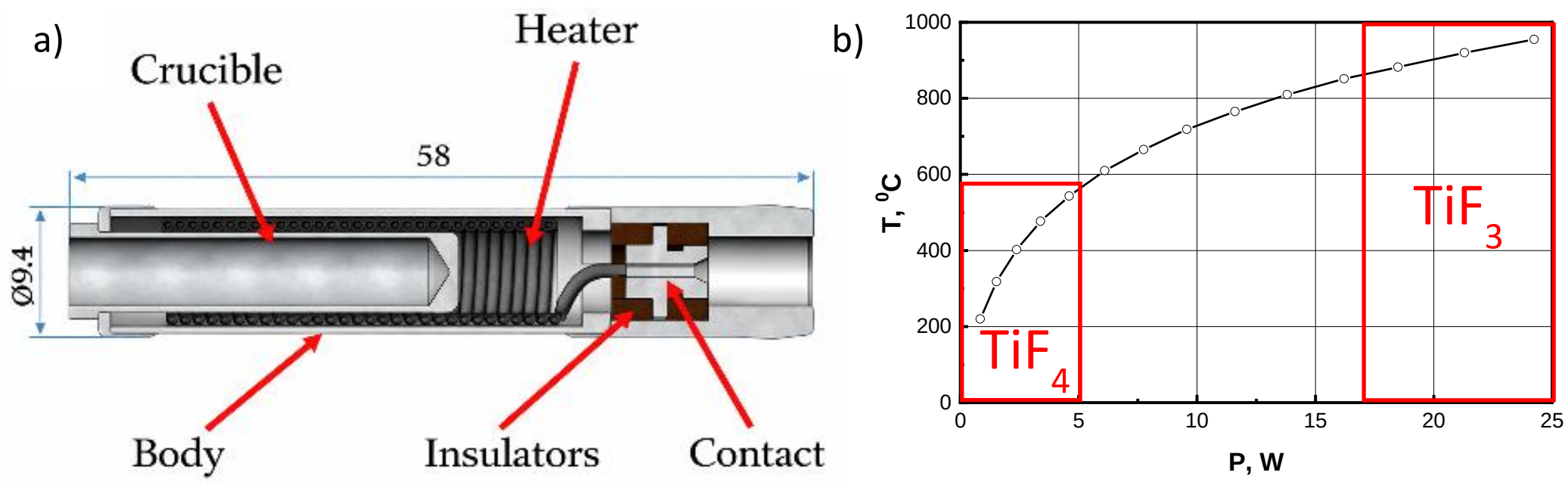

**Fig. 2.** – Low-temperature resistive oven (a). The crucible temperature as a function of the heating power (b).

When injecting $TiF_3$ molecules into the source, high currents of titanium ions were extracted when running the source with helium as the support gas, injected microwave power of 300 W and the oven power of around 18 W. The extracted ion current of $Ti^{9+}$ reaches the level of 60 μA (Fig.3a) in these optimized conditions. After tuning the source, a long-term stability test was performed for 24 hours without adjustments of any source parameters (Fig.3b). During the test period, the extracted $Ti^{9+}$ current decreased by 50%. After operator's intervention and re-tuning of the source by adjustment of the bias electrode voltage, magnetic field profile and the helium flux, the extracted currents were restored to their initial level even without increasing the heater power. Thus, the source stability is found to be adequate for the prolonged operation.

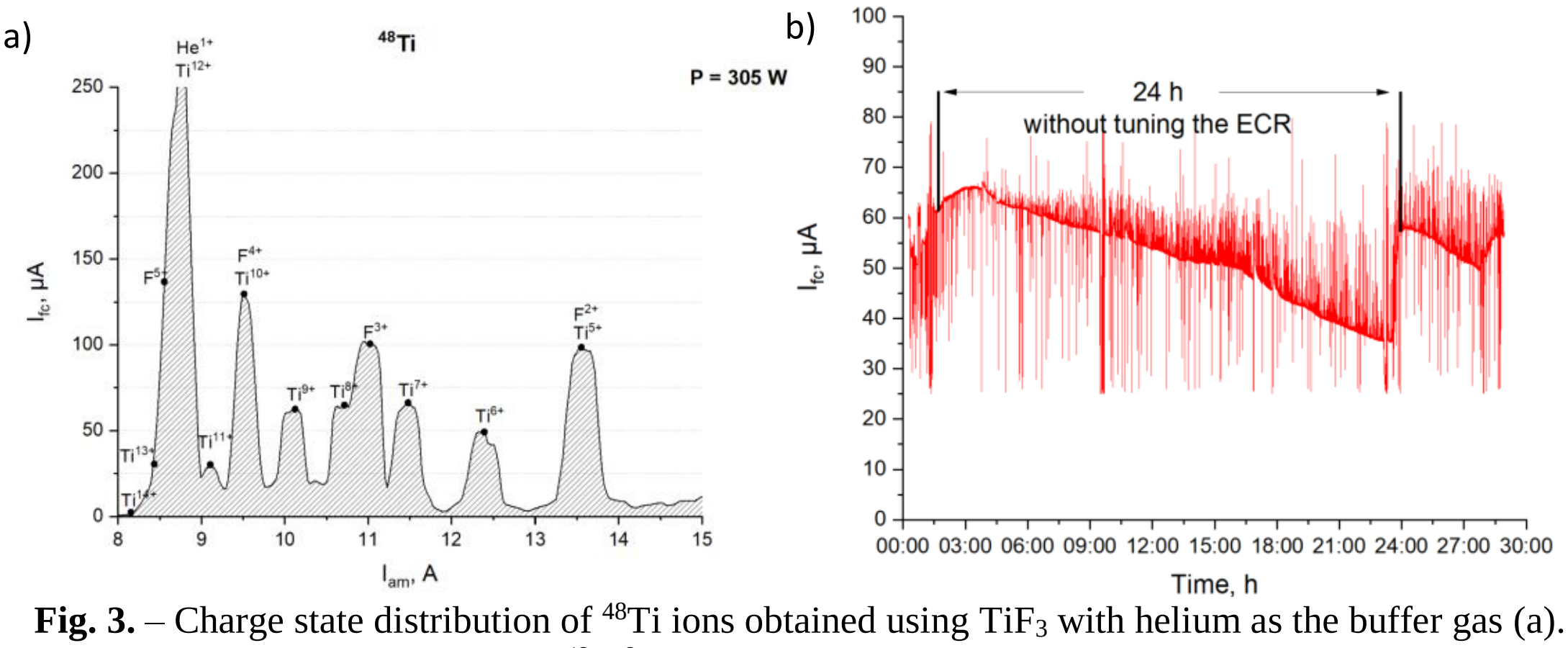

**Fig. 3.** – Charge state distribution of $^{48}Ti$ ions obtained using $TiF_3$ with helium as the buffer gas (a). Stability of the $^{48}Ti^{9+}$ ion beam current over time (b).

At the average $^{48}Ti^{9+}$ beam current of 60–65 μA, the $^{48}Ti$ consumption was 2.5 mg/h, with an overall ionization efficiency of 4.5%. Using the MIVOC method with an average $^{48}Ti^{9+}$ ion beam current of 40 μA, the $^{48}Ti$ consumption is 0.5 mg/h, and an overall ionization efficiency is 10%. Despite the higher material consumption and lower ionization efficiency when using $TiF_3$, it should be noted that this method allows increasing the overall intensity of the desired titanium ion beam by more than 30%.

When evaporating $TiF_4$ fluoride, no noticeable currents of the extracted titanium ion currents were detected, with only fluorine, hydrogen and oxygen ions in the spectra. The material is highly hygroscopic, and the release of oxygen during heating reduces the ionization efficiency and promotes secondary chemical reactions within the plasma chamber. Second, considering the low sublimation temperature of the material, a control of the evaporation process is challenging because of the additional heating caused by plasma and microwave power absorption in the crucible. Our best result with $TiF_4$ injection was 2 μA for $^{48}Ti^{11+}$ ion current only, which prohibits its further use.

## 4. Metallic titanium etching by $SF_6$ plasma

Injection of titanium atoms into the plasma under the fluorine etching was investigated by inserting the metallic titanium into the source chamber and igniting the ECRIS plasma under injection of sulfur hexafluoride as the main gas and helium as the support gas. In the first experiment, a 100 μm thick titanium foil was placed inside a 100 μm tantalum screen covering the radial walls and at the end cap close to the plasma electrode. The injection flange was left open and we used a stainless-steel biased electrode. The screen was thermally isolated from the walls to promote evaporation from the hot surface of the titanium fluorides created under impact of fluorine ions and radicals. The titanium foil inside the screen after running the source with $SF_6$ for approximately 24 hours is shown in Fig. 4. The radial parts of the foil were heavily damaged at the places where plasma touches the wall, while at the end cap the foil remains relatively intact.

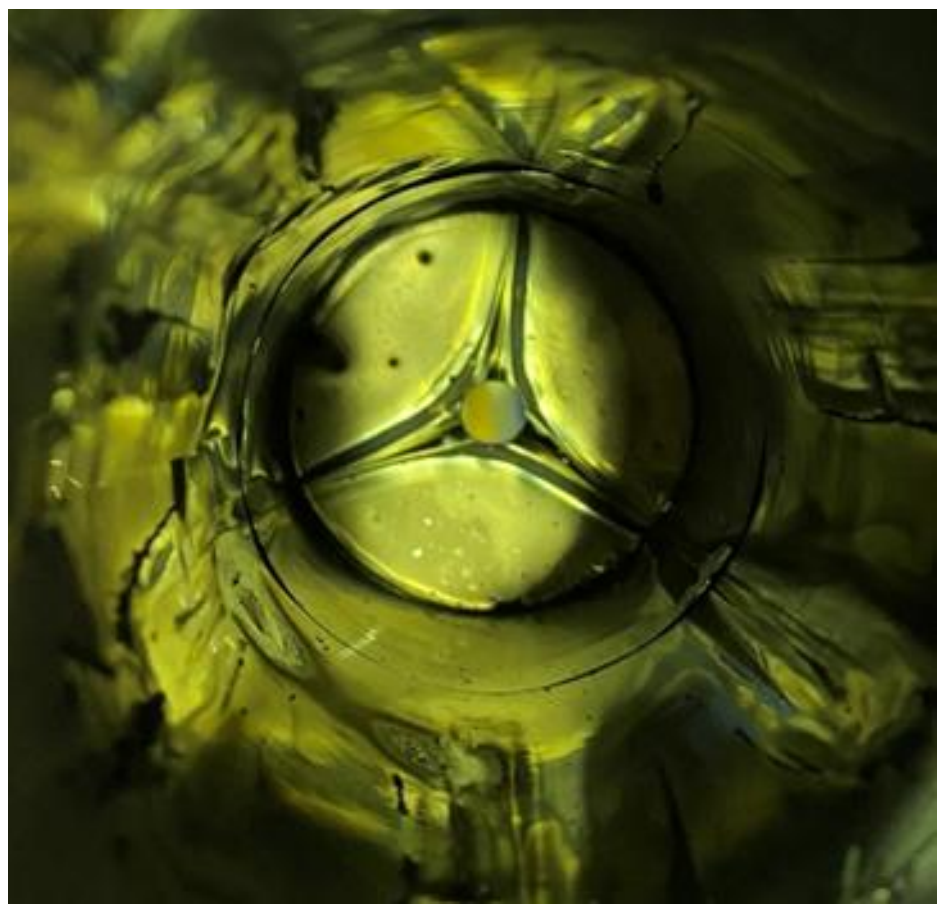

**Fig.4.** – Titanium foil after operation.

In this configuration, intense and stable beams of titanium ions were produced. The extracted beam spectrum is shown in Fig.5a, with the $^{48}Ti^{11+}$ ion current of 70 μA. The results of the long-term stability test are shown in Fig.5b, where the accelerated beam current of $^{48}Ti^{10+}$ ions (red, right scale) at exit of DC-280 cyclotron is shown, as well as the total leak current of the source (black, left scale) for the run duration of 11 hours.

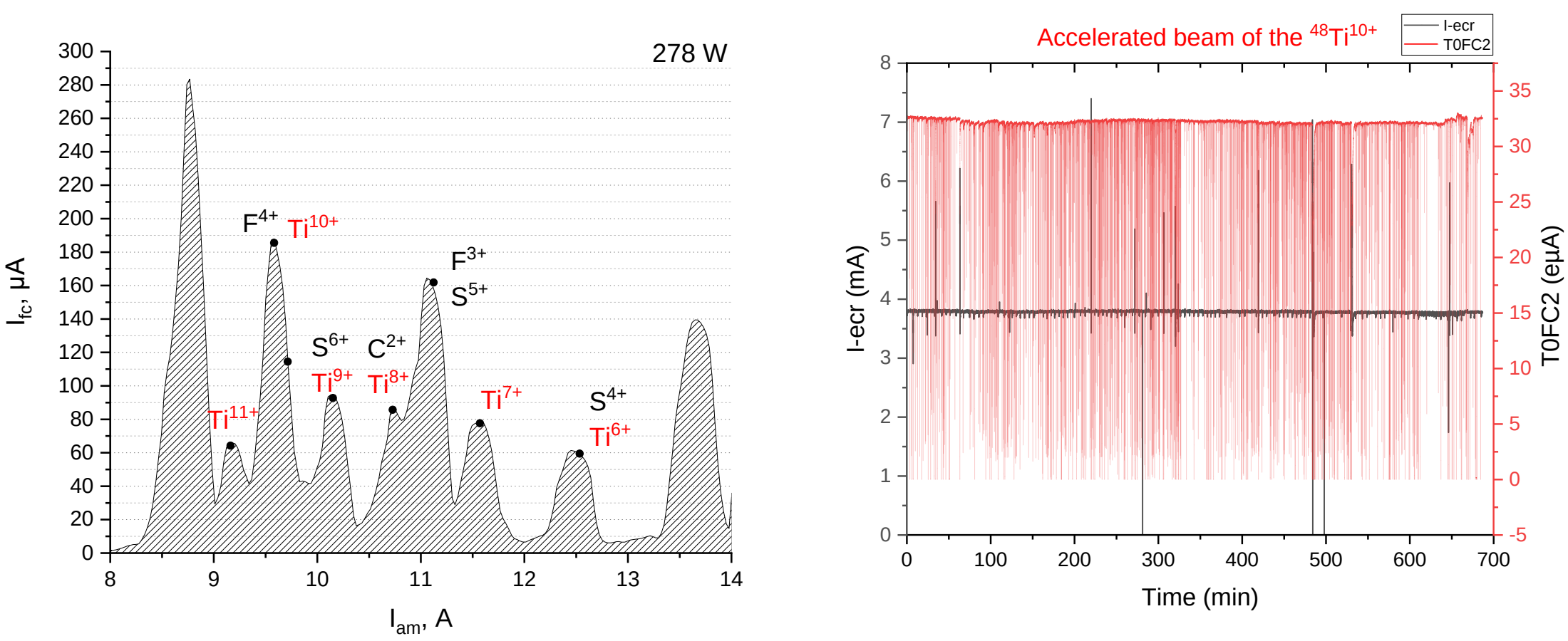

**Fig. 5.** – (a) Charge-state distribution of titanium ions; (b) $^{48}Ti^{10+}$ ion current as a function of time (red, right scale) and total leak current of the source (black, left scale).

In attempt to find the optimal titanium localization inside the source, the foil was removed from the source and the titanium biased electrode (D = 30 mm) was mounted on the injection axis. Compared to the previous configuration, the effective surface area of plasma contact with titanium was reduced by more than a factor of 20, amounting to 250 $mm^2$. The extracted ion current decreased by a factor of two, reaching 30 μA for $^{48}Ti^{11+}$ ions.

## 5. Conclusion

Three alternative approaches for the production of titanium ions in the DECRIS-PM ECR ion source were investigated in this work: direct evaporation of $TiF_3$ and $TiF_4$ fluorides, and the generation of volatile titanium species in the plasma via the reaction of titanium foil with $SF_6$. It was shown that $TiF_3$ allows the production of stable medium-charge-state titanium ion beams with currents up to 60–65 μA, although a gradual decrease in intensity over time was observed due to decomposition of the material in the crucible.

Evaporation of $TiF_4$ proved to be limited due to its high hygroscopicity and unstable behavior upon heating, resulting in low beam currents (up to 2 μA for $Ti^{11+}$). The most promising method is the use of titanium foil with $SF_6$. In this case, a 70 μA $Ti^{11+}$ ion beam was obtained with confirmed long-term stability during acceleration experiments, without the need for adjustments to the ECR source during operation.

The application of fluorides and fluoride-forming precursors represents an effective alternative to conventional methods for titanium injection into ECR ion sources. Future work should focus on optimizing the evaporation systems, identifying the optimal plasma-exposed titanium surface area, and protecting key components of the ECR source against corrosion during prolonged operation. The results provide a basis for applying these fluoride-based methods to rare titanium isotopes (in particular $^{50}Ti$), potentially expanding the experimental capabilities in nuclear physics and super-heavy element research.

## Author statement

**D.K. Pugachev:** Conceptualization, Methodology, Investigation, Formal analysis, Writing – original draft, Project administrator, Supervision. **K.I. Berestov:** Investigation, Resources. **A.E. Bondarchenko:** Investigation, Resources, Formal analysis. **V.N. Loginov:** Investigation, Resources. **A.N. Lebedev:** Investigation. **E.Yu. Kidanova:** Writing – review & editing. **K.I. Kuzmenkov:** Investigation, Resources. **V.E. Mironov:** Conceptualization, Methodology, Investigation, Writing – review & editing. **D.S. Podoinikov:** Methodology, Investigation, Validation, Data curation, Visualization. **D. S.Yakovlev:** Methodology, Investigation, Validation.

## Declaration of competing interest

The authors declare that they have no known competing financial interests or personal relationships that could have appeared to influence the work reported in this paper.

## Data availability

Data will be made available on request.